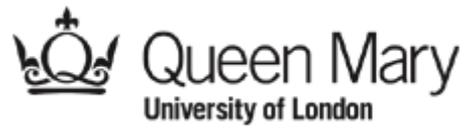

Queen Mary Law Research
Paper No. 370/2021

# Non-Asimov Explanations Regulating AI through Transparency

**Chris Reed, Keri Grieman, and Joseph Early**

# Non-asimov explanations

# regulating AI through transparency

*Chris Reed,\* Keri Grieman,\*\* and Joseph Early\*\*\**

**ABSTRACT**

*An important part of law and regulation is demanding explanations for actual and potential failures. We ask questions like: What happened (or might happen) to cause this failure? And why did (or might) it happen? These are disguised normative questions – they really ask what ought to have happened, and how the humans involved ought to have behaved.*

*If we ask the same questions about AI systems we run into two difficulties. The first is what might be described as the 'black box' problem, which lawyers have begun to investigate. Some modern AI systems are highly complex, so that even their makers might be unable to understand their workings fully, and thus answer the what and why questions. Technologists are beginning to work on this problem, aiming to use technology to explain the workings of autonomous systems more effectively, and also to produce autonomous systems which are easier to explain.*

*But the second difficulty is so far underexplored, and is a more important one for law and regulation. This is that the kinds of explanation required by law and regulation are not, at least at first sight, the kinds of explanation which AI systems can currently provide.*

*To answer the normative questions, law and regulation seeks a narrative explanation, a story. Humans usually explain their*

\* Chris Reed, Professor of Electronic Commerce Law at the Centre for Commercial Law Studies, Queen Mary University of London

\*\* Keri Grieman, doctoral researcher and research associate, The Alan Turing Institute and Queen Mary University of London, University of Oxford

\*\*\* Joseph Early, PhD Student at the AIC Research Group, University of Southampton

*decisions and actions in narrative form (even if the work of psychologists and neuroscientists tells us that some of the explanations are devised ex post, and may not accurately reflect what went on in the human mind). At present, we seek these kinds of narrative explanation from AI technology, because as humans we seek to understand technology's working through constructing a story to explain it. Our cultural history makes this inevitable – authors like Asimov, writing narratives about future AI technologies like intelligent robots, have told us that they act in ways explainable by the narrative logic which we use to explain human actions and so they can also be explained to us in those terms. This is, at least currently, not true.*

*This chapter argues that we can only solve this problem by working from both sides. Technologists will need to find ways to tell us stories which law and regulation can use. But law and regulation will also need to accept different kinds of narratives, which tell stories about fundamental legal and regulatory concepts like fairness and reasonableness that are different from those we are used to.*

*November 2021*

**TABLE OF CONTENTS (STYLE: TOC H CLP)**

# Table of Contents

# 1. INTRODUCTION

Non-lawyers think that all law consists of rules, but lawyers know that much of it is a series of questions. This is particularly so when a legal system decides to regulate something, or when we are attempting to decide if some defect or failure should give rise to legal liability.

There are two main questions which we ask here:

- *What?* What ought to happen? What did happen? What should have happened?

- *Why?* Why will it happen? Why did it happen? Why wasn't it prevented?

These questions have served us very well when regulating human actions and deciding on liability where those actions cause loss or damage. But they work less well if we remove the humans from the loop[1] and instead hand over the decision-making and resulting actions to AI systems.

One reason for this difficulty is that these questions are primarily normative, not factual. The most important aspects of their answers, for law and regulation, tell us about how events *ought to* have occurred compared to how they actually did. When we ask them of the humans who made decisions and initiated actions, we are trying to find out if those humans acted properly. We, or more accurately law and regulation, have over the years established standards for proper human behaviour. We know how humans ought to have behaved. But we are far less sure how AI systems ought to behave.

---

[1] An AI system which is 'human in the loop' makes a recommendation to a human, but the ultimate decision is still left to that human. The traditional questions asked by law and regulation can thus be applied to that human decision. The oft-expressed fear that humans will automatically assume that a computer's advice is more credible than their own judgment seems, according to empirical research, to be a myth – BJ Fogg & Hsiang Teng, 'The elements of computer credibility' (1999) CHI '99: Proceedings of the SIGCHI conference on Human Factors in Computing Systems May 1999, 80, 81.

As an example, take the well-known Tesla crash in the US in 2016. It appears that an important cause of the crash was that the autonomous driving technology misidentified another vehicle as being part of the sky, and so did not brake or turn to avoid collision.[2] No human driver ought to make such a mistake, or rather, no human driver ought to make such a mistake for this reason. And yet, up until this crash, Tesla cars had driven themselves on the roads with far fewer accidents of any kind than would have been caused if they had been driven by humans. On one measure, the technology performs worse than humans; on a different measure, it performs much better. Which is the correct standard? Or is it neither?

For liability, this problem is one which time could solve. Through several hundreds of decisions about liability for crashes involving autonomous vehicles, the courts of each country would be likely to evolve suitable standards of performance for AI systems. Admittedly, we might not wish to live with the uncertainty until this evolution is complete, and there would still be uncertainty about how well newly developed AI systems met those standards, or whether the standards should later evolve to reflect improvements in AI design.

Regulators cannot wait that long. Their job is to devise regulations which mitigate the risks to society caused by the activities they regulate.[3] This requires them to set some standards in

---

[2] Larry Greenemeier, 'Driverless Cars Will Face Moral Dilemmas', *Scientific American* 23 June 2016, http://www.scientificamerican.com/article/driverless-cars-will-face-moral-dilemmas/; Tesla Motors statement, 30 June 2016 – https://www.teslamotors.com/en_GB/blog/tragic-loss.

[3] In some cases, the mitigation might be through prohibiting the use of AI for a particular purpose – see, e.g., the list of prohibited AI practices in Article 5 of the proposed EU Artificial Intelligence Act (Proposal for a Regulation of the European Parliament and of the Council Laying Down Harmonised Rules on Artificial Intelligence (Artificial Intelligence Act), COM(2021) 206 final 21 April 2021). Explaining the decision-making of such an AI would not alter the prohibition, and therefore such AI systems fall outside the scope of this chapter.

That said, a decision-making explanation might be useful in deciding if the AI falls within a prohibition. For example, Art 5(1)(b) of the proposed AI Act prohibits use of AI which exploits vulnerable persons, and whether or not such exploitation was occurring might not be knowable if the AI's decision-making cannot be explained.

advance, rather than waiting for the risks to eventuate and then deciding retrospectively what should have happened instead.

## 2. EXPLANATIONS

In order to do their jobs, courts and regulators need answers to the *what* and *why* questions. In a world of human decision-makers these answers come in the form of explanations.

Let us suppose that a doctor misdiagnoses a patient's condition. On its own, this tells us nothing about whether the doctor failed to meet a normative obligation. Even if all the necessary standards are met, some medical diagnoses will be wrong. So instead, we interrogate the process through which the doctor made the diagnosis: what information did she take into account or ignore, and what were her thought processes when deciding what diagnosis to give based on the information she considered relevant? That explanation is given in a narrative form – it is the *story* of how the doctor undertook the diagnosis.

Now let us suppose that an AI system is undertaking the diagnosis. The obvious course of action is for a regulator or a court to demand a similar explanation, a story about the AI's decision-making processes. Such an explanation is the most important element of transparency, which has been recommended as one of the main tools for AI regulation.[4] The aim is that the AI, or its developers, should be able to explain the decision-making options available to the technology in each case, and the choices it made between them. If achievable, this would help resolve responsibility and liability

---

[4] See European Parliament resolution of 20 October 2020 with recommendations to the Commission on a framework of ethical aspects of artificial intelligence, robotics and related technologies (2020/2012(INL)), which calls for explainability in addition to transparency.

See also European Commission, *White Paper On Artificial Intelligence – A European approach to excellence and trust*, COM(2020) 65 final 19 February 2020; EU High-Level Expert Group on Artificial Intelligence, *Ethics Guidelines for Trustworthy AI*, 8 April 2019; UK House of Commons Science and Technology Committee (2016) *Robotics and artificial intelligence* (HC 145 12); US Department of Transportation/NHTSA (2016) *Federal Automated Vehicles Policy – Accelerating the Next Revolution in Road Safety*.

questions and assure regulators that the AI will not cause unexpected individual or social harm.

The question is whether the kind of narrative explanation that regulators and courts expect is actually achievable for AI.

## 2.1 EXPLANATIONS THROUGH METRICS

The easiest explanation which can be offered about an AI's decisions is given in numerical form, setting out how the AI has performed according to some chosen metric. Thus, the developers of a facial recognition AI might demonstrate that it can recognise faces it has previously 'seen' with 95% accuracy, or an autonomous vehicle might be shown to have 80% fewer accidents per 10,000 kilometres than human drivers do on average. This tells us something about how well the AI performs its task overall, but little or nothing about *how* it does so in each individual case. These metrics can also be misleading if the data are unrepresentative of the real-world cases in which the AI system could be used.

An additional issue with metrics is that there are multiple measures of performance which could be chosen when developing the AI. Optimising its performance against a particular metric may not optimise for the other metrics which could have been chosen. Our autonomous vehicle might have fewer accidents than human drivers, but more fatal accidents, and this might not be a better outcome overall. Therefore, an important facet of explainability lies in choosing an appropriate metric to evaluate the performance of an AI system.[5] And metrics are always a proxy for what we *really* want to assess; in this case whether the autonomous vehicle is safe enough for use on the road.

Further, there might be multiple AI solutions to a problem which score differently on the chosen metric, but one of those lower scoring solutions could still be preferable to the other choices. For example, a disaster response robot could choose a longer path to reach its objective as it avoids going through a weakened building that might collapse – something that is worse if measuring time to objective or fuel consumed, but is better when measured against the risk of damage to the robot and the likelihood of completing the mission.

---

[5] Maria Fox, Derek Long, and Daniele Magazzeni. 'Explainable planning'. *arXiv preprint arXiv:1709.10256* (2017).

From a regulatory standpoint, the choice of metrics used when optimising and testing an AI is an important issue. But it should by now be clear that metrics alone are not enough to satisfy the explanatory demands of law and regulation. Something closer to how humans explain their actions will be needed.

**2.2 ASIMOV EXPLANATIONS**

It is worth repeating here the questions set out in section **Error! Reference source not found.** which law and regulation ask about decision-making:

- *What?* What ought to happen? What did happen? What should have happened?

- *Why?* Why will it happen? Why did it happen? Why wasn't it prevented?

When humans are being regulated, we seek answers in the form of a narrative, explaining how the human went about making the decision in question. Then we can compare this answer to our chosen standard of human behaviour, such as taking reasonable care.

If we seek similar explanations about how an AI made its decisions, we are asking for what we, the authors, will call 'Asimov explanations'.

Stories of intelligent machines have been with us for millennia.[6] In *Politics*, Aristotle wrote:

… if every instrument could accomplish its own work, obeying or anticipating the will of others, like the statues of Daedalus, or the tripods of Hephaestus … chief workmen would not want servants, nor masters slaves.[7]

---

[6] For a helpful overview of the earliest stories, see A History of Artificial Intelligence: Antiquity, https://ahistoryofai.com/antiquity/.

[7] Aristotle, *Politics* (trans Benjamin Jowett, Oxford: Clarendon Press 1885) vol 1, 6; Book I part IV.

Around a thousand years later, and about a thousand years ago, the Indian story book *Śṛṅgāramañjarīkathā* told of King Bhoja's pleasure garden which contained a doll who could speak, along with a range of other automata.[8]

But the most influential stories about intelligent machines are undoubtedly those of Isaac Asimov, who published stories on this topic in the 1940s in the magazines *Super Science Stories* and *Astounding Science Fiction*, and then published them in book form as *I, Robot* in 1950.[9] In these stories, intelligent robots are constrained to obey the three laws of robotics[10] that Asimov invented. The stories explore the logical contradictions between these laws, which result in the robots behaving very differently from what was expected.

The importance of these stories is that the decisions and actions of the robots are explained to humans in terms of human logic. Observers of the robots induce their 'reasoning' and explain it using human language. These explanations are given as a narrative of the robots' 'thought' processes, and explain those processes just as a human actor might explain their own actions or decisions (or more accurately, as a human acting solely in accordance with a set of rules might do). Asimov's stories contain internal stories about how robots think, and they tell us that robot thinking can be explained via telling stories.

This cultural understanding that intelligent machines can be explained via stories has led to proposals to regulate AI by demanding narrative explanations about how it makes decisions[11], or even the imposition of express regulatory obligations to produce such explanations.[12] These demands, expressed through law and regulation, are based on a belief that such explanations are possible. But our human beliefs about what is possible (apart from the beliefs of those who have studied AI technology closely) are culturally derived, originating in fictional narratives rather than scientific papers. They are likely to be wrong.

---

[8] See Daud Ali, 'Bhoja's Mechanical Garden: Translating Wonder Across The Indian Ocean, Circa 800–1100 CE' (2016) 55 History of Religions 460, 462–3. The article later discusses other depictions of automata, many of which act autonomously, in Indian stories of that period.

[9] Isaac Asimov, *I, Robot* (Gnome Press 1950).

[10] 1. A robot may not injure a human being or, through inaction, allow a human being to come to harm.

2. A robot must obey orders given it by human beings except where such orders would conflict with the First Law.

3. A robot must protect its own existence as long as such protection does not conflict with the First or Second Law.

[11] *Draft Report with recommendations to the Commission on Civil Law Rules on Robotics* (2015/2103(INL), European Parliament Committee on Legal Affairs 31 May 2016).

As we will see, AI cannot currently be explained in this way, and might never be able to explain itself solely by means of stories. This chapter therefore needs to investigate what kinds of explanations *can* be given.

## 3. THE BLACK BOX PROBLEM

Technical systems whose workings are not understandable by humans are often described as 'black box' systems. Some AI systems are not black boxes in this sense – for ones that use simpler mechanisms, it is possible to accurately describe the processes through which the AI reached its decision. Such a system is inherently interpretable, and an interpretation of a decision is a full, though highly technical, explanation of how that decision was arrived at.

But from the perspective of law and regulation, a technical interpretation might be equally as opaque as a true black box system. The relevant question, from that perspective, is whether the person who is entitled to ask the question can understand the explanation. If not, the AI is functionally a black box in this context, even if in some other context (AI development, for example) the explanation might be comprehensible. For example, the developer of a machine learning-based AI might be able to explain to another AI developer how and why the AI reaches its decisions, but that explanation tells the user of the AI nothing. All that the user knows is that he is ignorant of the AI's workings, and that it is *de facto* a 'black box'.

---

[12] See e.g., *Federal Automated Vehicles Policy – Accelerating the Next Revolution in Road Safety* (US Department of Transportation/NHTSA, September 2016). Article 13(2)(f) of the EU General Data Protection Regulation, Regulation 2016/679, entitles data subjects to 'meaningful information about the logic involved' in automated decision-making involving their personal data. The proposed EU Artificial Intelligence Act (n 3) adopts a more nuanced approach in article 13(1): 'High-risk AI systems shall be designed and developed in such a way to ensure that their operation is sufficiently transparent to enable users to interpret the system's output and use it appropriately.' See also Article 14(4)(c) requiring that those responsible for high-risk AI systems should be able to correctly interpret their outputs.

In this sense, the opacity of the AI system also depends on when and how the question is asked. If the AI has produced a result which causes loss or damage, it may be possible to obtain some kind of answer depending on what type of AI system it is. However, explanation in advance, to help a regulator decide if an AI meets any requirements necessary for its use, is more difficult. In terms of their capacity to have their decision-making explained, AIs can be classified into two types.

*Rule-based* AI technologies implement sets of rules (analogous to IF … THEN … statements), and these sets of rules result in a decision tree. In theory, these rules could be hand-crafted, and the person doing so could therefore explain the decision-making process in terms which a human might understand. Each decision by the AI is the result of a single path through the decision tree to the output, and that path could be described as the 'reasoning' which led to its decision. However, all but the simplest rule-based AIs are likely to generate their rule sets through machine learning processes, such as genetic techniques which combine parts of two current rule sets and keep the 'offspring' which perform better than their parents. The resulting rule set is thus not an implementation of the reasoning processes of a human mind. If it were subsequently analysed by a human, some description of its reasoning for an individual decision could be produced, but that description will be of complex and technological reasoning, and unlikely to produce the kind of narrative explanation that non-technologists understand. Stories about human decision-making concentrate on motivation and intention, neither of which will be present here. There is also a likelihood that the logic of the resulting rule set may well be too different, detailed and complicated for the human mind to understand fully, what Burrell describes as:

> opacity that stems from the mismatch between mathematical optimization in high-dimensionality characteristic of machine learning and the demands of human-scale reasoning and styles of semantic interpretation.[13]

*Pattern-matching* AI technologies such as neural networks do not make decisions by following a path through a decision tree. They identify and match patterns in their inputs, and from those patterns they induce (rather than deduce) their output.[14] These systems are highly probabilistic – the output of an image recognition AI would not be 'this picture is of a moose' but rather 'this picture is more likely of a moose than any other animal' (possibly with a probability value for that likelihood). The AI learns how to make its decisions by analysing a large and comprehensive training dataset, and is then tested against a substantial real-world dataset. This process is iterated until the AI succeeds on real-world data sufficiently well to be put into use. From a non-technologist perspective, it 'just knows'. This makes it difficult to explain how the technology came to its decision, and thus how any loss or damage was caused. It is likely to be near-impossible to explain it in narrative terms.[15] Even if a rule set approximating the AI's decision-making could be reverse engineered, those rules might not convey anything meaningful to humans – 'IF pixel at address X,Y has colour value > N THEN …'.

For both technologies, after-the-event explanations are often possible, although they may only be properly comprehensible to a few, highly-qualified humans. What, though, of explanations in advance, before the AI system is put to use? Regulators, and others such as insurers, might well want such explanations to assess the risks which arise from using the AI and how well they have been anticipated and guarded against. And the wider public might want such an explanation to persuade them to accept the technology – most citizens would be unconvinced by autonomous vehicles if all that they were told was, 'We can't explain how it works, but it's really safe.'

Generating an explanation in advance through human analysis of an AI's workings is particularly difficult. Algorithmic AIs are hard to explain because there are so many paths through the decision tree, maybe millions of paths in some cases. Small changes in inputs can result in very different outcomes. Explaining all these paths will not provide what is wanted – the human need for narrative requires an abstraction, a coherent collective story into which all these different paths fit. Such a narrative might not even exist; if it does, the human mind may not be up to the task of constructing it. For example, devising an advance narrative explanation of the workings of a neural network is a particularly intractable problem for human analysts, because there is no logic (in the human sense) behind its decisions.[16]

---

[13] Jenna Burrell, 'How the machine "thinks": Understanding opacity in machine learning algorithms' (2016) Big Data and Society 1, 2.

[14] In some cases, the human developer instructs the system what it should be looking for (supervised learning), in others the system just learns whatever it can (unsupervised learning).

[15] Burrell, n 13, 5–7.

All this suggests that human creators of AI will rarely be able to provide the narrative explanations which law and regulation currently demand. This is a problem, because demanding narratives as a precondition for allowing use of an AI (or granting insurance, which is a precondition of use if the AI producer wishes to avoid insolvency) will in many instances amount to prohibition on using that AI at all.

## 4. TECHNOLOGY TOOLS FOR EXPLANATION

So can technology help us to produce the explanations we want for law and regulation? There are two parts to this question. The first is what technology can actually tell us about the decision-making processes of AIs, both in advance and after the event. The second is how we can fit that information into our legal and regulatory explanation-demanding systems.

Answers to the first part are likely to come from the fast-developing field of eXplainable AI (XAI). The goal of XAI is to design tools that can provide explanations for the decisions of complex autonomous systems. The purpose of these explanations is to assist humans to understand the decision-making process, focusing on a number of key drivers. These include confidence, trust, safety, ethics and fairness.[17] By exposing the reasoning of an AI system, XAI can lead to improved performance in future iterations.[18]

---

[16] Humans are happy with making illogical decisions, of course. The music or food one likes is not decided through logical processes. But these kinds of decisions are deliberately excluded from the sphere of law and regulation. Where an activity falls within the legal and regulatory sphere, humans are expected to give narrative and logical explanations of their actions. The explanatory logic used in law and regulation tends to be simple propositional logic, for example: 'IF it is snowing THEN drive slower'.

## 4.1 XAI Techniques

Developments in XAI are advancing rapidly, and there is as yet no consistent terminology or taxonomy of XAI techniques. However, a recent survey of the field[19] suggests that the following categories of XAI research might usefully group related techniques together:

1. Saliency techniques. These identify the relative importance of different inputs to the AI in producing particular outputs – for example, the regions of tissue that contain cancerous cells. Results are often presented visually or quasi-visually (e.g., in the form of a heat map of words or phrases for textual analysis AIs). The idea here is that these representations will produce patterns which humans can map to their own understandings of how decisions in that field are made, and thus use them to explain the AI's decision-making.

2. Signal methods. These are used for image recognition neural networks, and identify how input images affect the values of

---

[17] Doran, Derek, Sarah Schulz, and Tarek R. Besold. 'What does explainable AI really mean? A new conceptualization of perspectives'. *arXiv preprint arXiv:1710.00794* (2017).

[18] Anjomshoae, Sule, et al. 'Explainable agents and robots: Results from a systematic literature review'. *18th International Conference on Autonomous Agents and Multiagent Systems (AAMAS 2019), Montreal, Canada, May 13–17, 2019.* International Foundation for Autonomous Agents and Multiagent Systems, 2019.

[19] Tjoa, Erico, and Cuntai Guan. 'A survey on explainable artificial intelligence (XAI): towards medical XAI'. *arXiv preprint arXiv:1907.07374* (2019). For alternative taxonomies, see e.g., Biran, Or, and Courtenay Cotton. 'Explanation and justification in machine learning: A survey'. *IJCAI-17 workshop on explainable AI (XAI).* Vol. 8. No. 1. 2017; Guidotti, Riccardo, et al. 'A survey of methods for explaining black box models'. *ACM computing surveys (CSUR)* 51.5 (2018): 1–42.

the neurons in a layer of the network. What that layer 'sees' can then be reconstructed and compared by a human to the original image, to discover which parts of the input image are detected by each layer. From this a narrative might be constructed in the case of, say, facial recognition: 'First the AI identifies the eyes and nose, the next layer finds the edges of the face, the third layer …'.

3. Verbal (or textual) interpretability methods attempt to translate symbolic processing into verbal 'IF … THEN …' rules. These methods are likely to be used on text analysis algorithms, because the input text can be used to construct the 'IF … THEN …' statements which explain the AI's decisions. In effect, these statements are a higher-level abstraction of the more complex set of rules actually embedded in the algorithm. One known problem with verbal interpretability is justifying the techniques used to produce the verbal 'IF … THEN …' statements – these techniques might still be 'black boxes' so far as the person receiving the explanation is concerned.

4. Mathematical modelling. This technique requires a mathematical model to be devised which matches (or perhaps more accurately: approximates) the relationship between inputs to the AI and its outputs. A technical expert will be able to understand that model, and it is hoped will also be able to explain it in non-mathematical terms to any human who requires an explanation. In effect, the human-incomprehensible workings of the AI are abstracted into a mathematical model which is understandable by some skilled humans, and those humans can explain them to other humans at an even higher level of abstraction.

5. Feature extraction (or importance). This identifies features in the input data (e.g., for medical diagnosis, the inputs relating to fitness, eating patterns and sleep patterns) and then identifies the features which are most strongly correlated for particular outputs and those which are not correlated. Feature extraction is thus a type of abstraction; it might find, for a particular disease, that when the AI makes its diagnoses, sleep and diet are closely correlated, whereas geographical residence and income are not. These correlations can be used for human explanations.

6. Sensitivity methods. These take individual decisions of the AI and make changes to its inputs, to see how they affect its outputs.[20] This can identify which inputs are most important for producing the decision. It can also offer a measure of reliability for the AI, because if tiny changes in inputs produce major changes in output, the AI might not produce reliable results on inputs it has not seen before. One difficulty with these techniques is generalising them to provide useful information about the workings of the AI overall, rather than just explaining individual decisions.

7. Optimisation (or decomposition). This attempts to find sub-elements of the AI which, for the same input data, produce outputs which are recognisably related to the full AI's output. This is a kind of abstraction of the AI, and the theory is that the abstraction can be interpreted (probably by technical experts) to discover information about the full AI's decision-making.

From these descriptions it is clear that there is no single tool which will be able to provide the explanation needed by law and regulation.[21] Different explanations are needed by different users of 'In the existing literature, "explanation" typically refers to an attempt to convey the internal state or logic of an algorithm. In contrast, counterfactuals describe a dependency on the external facts that led to that decision.' Sandra Wachter, Brent Mittelstadt and Chris Russell, Counterfactual Explanations Without Opening the Black Box: Automated Decisions and the GDPR (2018) 31 Harvard Journal of Law & Technology 841, 845.

---

[20] There is a growing legal literature on counterfactuals, which are a type of sensitivity method. Counterfactual explanations function by reiterating a data process with the smallest possible change to determine which parts of the data are influencing a decision. Small tweaks are made to the data, then the 'question' put to the AI is asked again and again, pinpointing which data points changed the outcome.

For a full discussion of counterfactual explanations, see Katja de Vries, Transparent Dreams (Are Made of This): Counterfactuals as Transparency Tools in ADM (2021) 8 Critical Analysis of Law 121.

See also Atoosa Kasirzadeh and Andrew Smart, The Use and Misuse of Counterfactuals in Ethical Machine Learning [2021] arXiv:2102.05085 [cs] <http://arxiv.org/abs/2102.05085> accessed 24 February 2021.

[21] Indeed, some tools are developed specifically to explain a particular AI's decisions, and thus would not be usable to explain other AIs.

explanations, for example the explainability requirements for a regulator or a developer would be different to those needed for an end user.[22] However, each tool potentially contributes something useful, and they might be used in combination to assist the explanation process.[23]

## 4.2 USING XAI TOOLS TO EXPLAIN

How these tools might be used depends very much on how well two factors are understood:

- The input data which the AI might receive; and

- The consequences for the external world which that AI's outputs might have. The range of decisions it can make will of course be known, but the potential consequences of those decisions might or might not be known, or even knowable.

### 4.1.1 INPUTS

For domains that have well-understood inputs, it is possible to have an understanding of how the system *should* work. This means that any explanations generated for an AI working in such a domain should match the expectations of humans who currently work in the domain. For example, in medical imaging the range of images which might be assessed is known, and doctors already know what they are looking for in those images. Thus, if they are provided with the explanations from an AI system, they can verify that the rules or techniques which the AI appears to have learnt match the image analysis rules which they apply themselves. An XAI explanation which highlighted the elements of an X-ray that leads the AI to a positive classification of cancer, for example, could be used by doctors to check whether these are the same elements which guide their own diagnoses.

However, in other domains we might not have a good understanding of the inputs, or the input space might be so large that the AI cannot be trained on every possible input it is likely to encounter. An autonomous vehicle used on Canadian roads might expect to encounter a moose or a bear, and thus be trained to recognise those animals, but a peacock would be as much of a surprise to that vehicle is it was to one of the authors when he encountered one on an English country lane.

---

[22] Sam Hepenstal and David McNeish. 'Explainable Artificial Intelligence: What Do You Need to Know?'. *International Conference on Human-Computer Interaction.* Springer, Cham, 2020.

[23] Langley, Pat, et al. 'Explainable agency for intelligent autonomous systems'. *Twenty-Ninth IAAI Conference.* 2017.

If some of its inputs are unknowable in advance, it is hard to say how an AI *should* work. Even if we can explain how it will behave if it encounters a moose or a bear, we can only guess what it will do when presented with a peacock. This does not mean, though, that XAI cannot provide some assistance, particularly in open-ended domains where the optimal strategy is unknown to humans. As an example, we can be left scratching our heads when an AI system outperforms us and we don't know how it makes its decisions. DeepMind's AlphaZero made radical and unexpected moves in the game of Go which ultimately proved to be beneficial later in the game, and expert players are still studying and analysing those moves.[24]

The explanations which XAI can provide can help in two ways here. First, they can expose some information about the AI's decision-making and thus provide some reassurance that it is not doing something untoward, such as making unlawfully biased decisions.[25] Second, they can increase our knowledge about a domain (e.g., by highlighting a previously unknown relationship or explaining why a particular course of action is beneficial). From a regulatory perspective this is helpful in ensuring the system aligns with long-term goals, such as improving industry standards, by revealing something new about how good performance can be achieved.

---

[24] Silver, David, et al. 'Mastering the game of Go without human knowledge'. *Nature* 550.7676 (2017): 354–9.

[25] Let us imagine an AI which selects students for a drama degree. Anecdotally, the culture of the acting profession has been welcoming towards those of a minority sexual orientation, which might attract such persons to attempt to enter the profession. Our AI, learning from previous applications and examples of accepted students, might therefore teach itself to rely on clues to sexual orientation in deciding which students to select. This would be unlawful, so an explanation sufficient to show it is unlikely to be doing this would be useful, even if that explanation cannot give a full picture of how the AI works. For a non-fictional example of unintended bias derived from AI training data, see Jeffrey Dastin, Amazon scraps secret AI recruiting tool that showed bias against women, Reuters 11 October 2018 (https://www.reuters.com/article/us-amazon-com-jobs-automation-insight-idUSKCN1MK08G).

### 4.2.2 CONSEQUENCES

When it is foreseeable that the outputs of a system might produce consequences which society will wish to avoid, such as deaths on the roads or inaccurate medical diagnoses, regulation attempts to ensure that these foreseeable failures do not occur. This entails putting the system in scenarios where a foreseeable fault could occur, and testing to see if it still acts as intended. XAI could assist in testing AIs by going beyond just observing the system's behaviour; it might allow the developers to ensure that the AI actually recognises the potential failure and takes steps to avoid it, rather than simply succeeding by some fluke occurrence. An example would be exposing an AI to adversarial examples designed to catch the system out, and seeing if it fails. If so, XAI tools will help in explaining why it fails so that developers understand how the system can be modified to avoid that failure in the future.

Unforeseeable consequences must be expected when it is impossible to test the AI system in every possible scenario it will ever encounter. The regulatory problem here is achieving sufficient reassurance that the AI will (or is at least likely to) act correctly in these circumstances, because the potential consequences of its decisions will by definition be unforeseeable.

If the internal decision-making of the system can to some extent be understood through explainability tools, and it is believed to fit well enough with known human decision-making in the same domain, a regulator might treat this as adequate assurance that if the AI encounters previously unseen situations it is likely to produce decisions whose effects are unlikely to be harmful (or at least, no more harmful than the effects of a human decision in such cases). XAI might even provide greater assurance than is achievable for human decision-making here, by finding the edge cases where the AI would act unexpectedly, exposing where there is a risk of unforeseeable consequences and perhaps even enabling the likely effects of the decision to be predicted. Requiring developers to use XAI tools to achieve a better understanding of how an AI makes decisions and how those decisions will affect the world around it might be a useful regulatory intervention for some types of AI.

# 5. RECONCILING EXPLANATIONS

So what does this tell us about what XAI can offer, working in conjunction with human AI developers, to explain AI to law and regulation? We could bring this all together, in broad terms, as follows.

a. The easiest explanation which can be given for an AI is some suitable metric about its performance. This might compare the AI's decision-making numerically to that of a human undertaking the same task, or it might explain what proportion of the cases it was tested on were decided correctly, as assessed by its human developers. These numbers are useful as one factor in deciding if the AI is sufficiently good at its task to grant it regulatory approval, if approval is needed, or to help decide if those producing or using it were in breach of their legal duties if a liability claim is made. However, the numbers only tell us about the overall performance of the AI – they give no clue about how well it decided in any individual case, or how it will perform in future cases.

b. XAI is sometimes able, in advance of an AI being put to use, to generate some information about the robustness and accuracy of the AI's decision-making. This will be by categorising some factors or reasons which are common to cases where the AI failed to make the decision which a human should have made, or which humans assess that the AI should have made. This might similarly be useful for deciding on regulatory approval or liability.

c. It can also be possible, in some instances, for XAI to identify which inputs most strongly influence the final decision and which have little effect. In advance of the AI making a decision, this will be an indication of which inputs are likely to be used in making a decision. After the event, it should be possible to say which inputs were or were not influential, though perhaps in terms of probabilities rather than certainties.

d. XAI might also be able to explain, to some extent, the order in which an AI builds up its decision, which could tell us something about dependencies. In the facial recognition example above, it might reveal that accurate identification of facial shape depends on accurate identification of eyes and nose, and so on. This could form the basis of a narrative explanation about how the AI is, or more accurately might be, working.

e. In the best case, from a legal and regulatory perspective, XAI might even produce an abstracted, high-level explanation of the *likely* 'reasoning' which a particular AI is using. But law and regulation will need to understand that this abstraction is a model of what the AI might be doing, which is developed from sample cases and the result of human interpretation of an XAI analysis of the AI's workings. The model might work only for some types of case, and not for others, so this is a dynamic explanation – over time the model might be disproved and an alternative model developed based on the XAI analysis, or XAI might improve the model so that its explanation is reasonably accurate for more cases. Because the model is both an abstraction and a simplification, it will not capture the full complexity of the decision-making, and thus cannot be relied on as a comprehensive explanation. Such a model is only the best guess that can currently be achieved about the AI's 'reasoning', a mixed product of machine analysis and human interpretation.

Working through a hypothetical example of an after-the-event explanation might be useful. Suppose that a fully autonomous vehicle collides with a pedestrian who has stepped into the road. What could an explanation aided by XAI look like compared to the explanation of a human driver?

A metrical explanation, or one which focuses on the general reliability and robustness of the AI (points a and b above), is of little help here. These are only useful in explaining whether it was safe to use the vehicle on the road at all, and we can assume that the fact that it was permitted on the road by regulators and insurers means that it had passed that test. So we might expect an explanation something like this:[26]

---

[26] This hypothetical explanation is loosely based on the Uber autonomous vehicle crash in Arizona, March 2018. See NTSB Preliminary Report HWY18MH010, https://www.ntsb.gov/investigations/AccidentReports/Reports/HWY18MH010-prelim.pdf; NTSB Board meeting documents, 19 November 2018, https://www.ntsb.gov/news/events/Pages/2019-HWY18MH010-BMG.aspx.

- Factual data from sensors about the speed the vehicle was driving, light conditions, etc.

- The AI driving technology identified that there was something in the road, but initially misidentified it as most likely (probability 0.82) a black plastic bag blowing in the wind and so did not slow down.

    o This happened because lidar[27] signals, used to identify 2D outline, colour etc, are processed faster than sonar signals, which contain supplementary information about the 3D shape of objects.

    o The obstacle recognition element of the AI identifies outline first, in this case as being probably that of a plastic bag.

- The AI then identified the obstacle as probably being a person (probability 0.91) using the additional data and braked, but there was insufficient time to stop before the collision.

    o As further data comes in, the nature of the obstacle is recalculated, adding revised lidar and sonar data as available.

    o Sonar data is more influential than lidar data in making the decision to brake (probability(sonar)*0.6 + probability(lidar)*0.4).

    o The model of the AI's reasoning suggests that assessment of the obstacle as more likely human than plastic bag using lidar data, and receipt (but not processing) of sonar data which would indicate its shape fitted human better than plastic bag, both happened at about the same time.

    o Braking started almost immediately thereafter (0.27 seconds).

---

[27] A technology commonly used for autonomous vehicles which uses the return signals from lasers to calculate distance from a target (here, the pedestrian) and also to create a 3D representation of the target.

The explanation of a human driver would be much briefer, something like this:

- I was driving below the speed limit and the light was poor, so I was keeping a good lookout.

- I saw the pedestrian in the road, but thought he was a black plastic rubbish bag blowing in the wind because his dark coat was flapping, so I didn't slow down immediately. In the circumstances, another human driver would have made the same misidentification.

- When I realised it was a pedestrian I braked hard, but this was too late to avoid the collision.

The first thing to note about these two explanations are that their main elements are broadly the same. However, for the AI, there is much more information about how it reached the various decisions it made.

The second thing is that the various explanations which XAI can provide about the AI driver do not form a coherent narrative about its 'motives' or 'intentions', which are an important part of the human driver's narrative. A human interpreter can take these XAI sub-explanations and weave them together to create something which approximates to such a narrative, but this is not the AI explaining itself – it is a human, generating an Asimov explanation of the AI's decision, based on observation of its workings by XAI tools.

The third thing is that the explanation given about the AI is largely probabilistic, except for the data about speed and light, which are objective. By contrast, the human driver's explanation is set out in definitive terms and is deterministic. Further thought should tell us, though, that the human driver's explanation is less reliable than that given by the AI. It depends on how accurately the driver can recall her speed, the thoughts which were going through her mind, and so on.

At first sight, these two explanations seem quite different. The explanation of the AI tells us what probably happened in decision-making, with reliable data to support those probabilities. The human explanation tells us definitively what is claimed to have happened, but without reliable data to support it.

Further thought should tell us that in fact the human explanation is also probabilistic. We cannot be sure that it is correct,

and so for legal purposes we have to make an assessment about how probable it is to be accurate. In a civil action, for example, we would ask whether the driver's version is more likely than not (probability 0.51 or greater) to be a true recollection.[28] Both explanations, AI and human, are uncertain. We might even argue that the main difference between them is that the AI explanation admits the uncertainty.

What, though, if we are asking a similar question in advance of there being an accident? That question might be whether the human driver, or the autonomous vehicle, will drive safely enough to be permitted onto the road at all.

For autonomous vehicles, this is where metrics and assurances about the reliability and robustness of the AI will come into play. Some measure of safety can be derived from comparative accident statistics about this AI's driving compared to that of human drivers, and is likely to be highly favourable to the AI or it would not be a commercially viable proposition. If a regulator wanted greater reassurance about particular driving situations where doubts had been raised, this might be provided by reviewing training failures as if they were real accidents and seeking the kinds of explanation set out above. If a generalised model of the AI's reasoning could be produced by XAI, the regulator could compare that to how humans are believed to make driving decisions in order to identify differences or gaps. Lastly, the AI developer's plans and processes to monitor and improve performance, particularly through analysing accidents, will be an important factor in deciding whether sufficient safety is likely.

Human drivers have it much easier. The majority of safety assurance is achieved through the training and examination required for a driving licence, and after that drivers are incentivised to continue to drive safely by criminal sanctions and legal liability, reflected in insurance premiums.

In both cases, the answer to the question is in fact a prediction, that the human or the AI will drive safely. For the human, that prediction is based on passing a driving test and the hope that the legal and financial incentives to drive safely will be effective. For the AI, there is likely to be more evidence on which to make the prediction, but as humans we find it hard to evaluate whether this evidence is more or less reliable or objective than the evidence underpinning our prediction for the human driver. Members of society, and regulators, are humans, and thus have an intuitive understanding about the reliability of predictions about other humans. AI reliability cannot be evaluated in the same way.

---

[28] Noting also, of course, the extensive body of psychological research which indicates that human memory can be distorted by belief. Thus, a driver who believes that he is a safe driver is likely, without intending to do so, to revise his memory of an accident to fit in with that belief. See further Rodriguez DN & Strange D, 'False memories for dissonance inducing events', (2015) 23(2) Memory 203.

## 6. CONCLUSION

As we have attempted to show in the previous section, a detailed analysis of the explanations which humans give for their decisions, and those which XAI might enable to be given about an AI, shows that they are likely to be much closer to each other than appears on the surface.[29] And yet our first instinct as lawyers and regulators is to accept the human explanations but reject the AI explanations as inadequate. Why might this be so?

Clearly it is the fault of Asimov and other tellers of fictional tales about intelligent machines. Humans explain themselves in definitive terms – this is how it happened, this is how I will decide – but AIs are predicated on uncertainty and only tell us probabilities – this is most likely to be how it happened, this is probably how future decisions will be made. XAI-assisted explanations for what has already happened might, as we have seen above, be similar enough to human explanations once their probabilistic nature is understood. Explanations about the future decisions an AI might make are, though, very different from those about human decision-making.

This clash of narrative expectations seems a plausible reason why we might demand more from an AI by way of explanation. But an AI whose future actions can readily be explained in deterministic terms – what *will* happen in its 'reasoning', not what is *probable* to happen – is likely to be much less able, and thus less useful, than the kinds of AI we have been discussing in this paper.

---

[29] Though we should note that this conclusion may not hold for all domains. As a simple example, an AI controlling a home heating system will be making very different decisions from a human controlling the same system manually, though their end aim (a comfortably warm home) is of course the same.

If we wish to secure the likely benefits from those kinds of AI, we will need to change our attitude to explanations. After all, the certainty which human explanations appear to offer is, we suggest, a false certainty. If we can accept that explanations for highly complex systems (including humans, who are highly complex) must inevitably be based on probabilities, we will have made a useful advance in law and regulation.